\documentclass[supp]{new_tlp}
\usepackage{supp}

\usepackage{comment}

\usepackage[T1]{fontenc}
\usepackage{amssymb}
\usepackage{graphicx}

\usepackage{amsmath}
\usepackage{textcomp}
\usepackage{listings}

\usepackage{graphicx, color}

\usepackage{ifthen}

\usepackage{fancyvrb}
\usepackage{amssymb}
\usepackage{hyperref}
\usepackage{enumitem}

\usepackage[scaled=0.70]{couriers}

\definecolor{gray}{rgb}{0.5,0.5,0.5}

\usepackage{url}

\urldef{\mailsa}\path|{sergio.castro,kim.mens}@uclouvain.be|    
\urldef{\mailsb}\path|pmoura@di.ubi.pt|

\usepackage{appendix}

\begin{document}

\title[Customisable Handling of Java References in Prolog Programs]
      {Customisable Handling of Java References\\in Prolog Programs}
%    {Customisable Handling of Foreign-Language References in Java-Prolog Programs}
%      {Customisable Management of Symbolic Foreign-Language References in Java-Prolog Programs}

\author[Sergio Castro, Kim Mens and Paulo Moura]
       {SERGIO CASTRO, KIM MENS and PAULO MOURA\thanks{This work is partially funded by ERDF through the CMPETE Programme and by FCT within project FCOMP-01-0124-FEDER-037281).}\\
       ICTEAM Institute, Universit\'{e} catholique de Louvain, Belgium \\
       CRACS \& INESC TEC, Faculty of Sciences, University of Porto\\
       %Rua do Campo Alegre, 1021/1055, 4169-007 Porto, Portugal\\ commenting out the address of Paulo's affiliation since we are not including our address
       \email{\{sergio.castro,kim.mens\}@uclouvain.be,pmoura@inescporto.pt}
      }

\date{foo}

%\titlerunning{A Software Ecosystem for Java--Prolog Interoperability}

%\author{
%Sergio Castro\inst{1}
%\and Kim Mens\inst{1}
%\and Paulo Moura\inst{2}}
%\and G\"unter Kniesel\inst{2}
%
%\authorrunning{A Software Ecosystem for Java--Prolog Interoperability}
%\authorrunning{Sergio Castro \and Kim Mens \and G\"unter Kniesel \and Paulo Moura}
%\authorrunning{Sergio Castro \and Kim Mens \and Paulo Moura}

%\institute{ICTEAM Institute,
%           Universit\'{e} catholique de Louvain, Belgium\\
%           \email{\{sergio.castro,kim.mens\}@uclouvain.be}
%           %\and 
%           %University of Bonn, Germany\\
%           %\email{gk@cs.uni-bonn.de}
%           Center for Research in Advanced Computing Systems, INESC--TEC, Portugal\\
%           \email{pmoura@inescporto.pt}
%}

%\institute{RELEASeD lab, ICTEAM Institute,\\
%Universit\'{e} Catholique de Louvain, Belgium\\
%\url{http://released.info.ucl.ac.be}\\
%\mailsb}
%\\
%\url{http://released.info.ucl.ac.be}}

\pubauthor{Sergio Castro, Kim Mens and Paulo Moura}

%\toctitle{A Software Ecosystem for Java--Prolog Interoperability}
%\toctitle{Authors' Instructions}
\maketitle

\begin{abstract}

Integration techniques for combining programs written in distinct language paradigms facilitate the implementation of specialised modules in the best language for their task.
In the case of Java-Prolog integration, a known problem is the proper representation of references to Java objects on the Prolog side.
To solve it adequately, multiple dimensions should be considered, including 
reference representation,
opacity of the representation,
identity preservation,
reference life span, and 
scope of the inter-language conversion policies.
This paper presents an approach that addresses all these dimensions, generalising and building on existing representation patterns of foreign references in Prolog, and taking inspiration from similar inter-language representation techniques found in other domains.
Our approach maximises portability by making few assumptions about the Prolog engine interacting with Java (e.g., embedded or executed as an external process).
We validate our work by extending \textsc{JPC}, an open-source integration library, with features supporting our approach.
Our \textsc{JPC} library is currently compatible with three different open source Prolog engines (\textsc{SWI}, \textsc{YAP} and \textsc{XSB}) by means of drivers.
\end{abstract}

%motivation, problem statement, approach, results, and conclusions. 

\begin{keywords}
Multi-Paradigm Programming, Language Interoperability, Logic Programming, Object-Oriented Programming, Prolog, Java
\end{keywords}

\vspace{-1em}

\section{Introduction}
\label{introduction}

Writing program modules in the language best suited for their task can greatly facilitate their implementation \cite{Mernik:2005:DDL:1118890.1118892}.
However, integrating modules written in different languages is not trivial when such languages belong to different paradigms \cite{Gybels2003}.
This is especially the case for Prolog programs integrated with an object-oriented language such as Java \cite{Denti2005217}.
One of the main problems of this integration is the proper representation of foreign language artefacts in the logic language, such that they can be conveniently manipulated and interpreted \cite{Gybels2003}. %\cite{wuy01b}.

The scope of this work concerns a portable approach to simplify the management and representation  %of references from an object-oriented language (Java) in a logic language (Prolog).
of Java object references in Prolog.
Studying existing solutions to this problem in Prolog, similar logic languages (e.g., \textsc{Soul} \cite{Roover:2011uq}) and even inter-language conversion libraries in other domains (e.g., Google's \textsc{Gson} library \cite{gson}), we have identified the following dimensions to be tackled:
1) reference representation; 
2) opacity of the representation; 
3) identity preservation; 
4) reference life span and 
5) scope of the inter-language conversion policies.
To maximise portability, our approach does not make any simplifying assumption regarding the architecture of the Prolog engine (e.g., such as it being embedded in the JVM). %, scenario where many of the reference management problems can be easily solved).
We validate our work by extending our \textsc{Java Prolog Connectivity} (\textsc{JPC})\footnote{\url{https://github.com/java-prolog-connectivity}} integration library \cite{2013-wasdett-castro}
with customisable support for managing Java references in Prolog. %TODO replace URL by https://java-prolog-connectivity.github.com/ if the site is ready before submitting the paper.
%Our library generalises and builds upon existing representation patterns of foreign references in logic languages.

This paper is structured as follows.
%Section \ref{conceptualProblem} 
Section 2 discusses the main Java reference representation issues in Prolog.
Section \ref{architecture} presents an overview of \textsc{JPC}'s architecture.
\textsc{JPC}'s approach for custom management of Java references in Prolog is discussed in section \ref{jpc}.
Section \ref{relatedWork} discusses related work.
Section \ref{conclusions} summarizes our conclusions and future work.
\vspace{-1em}

\section{The Problem of Representing Java References in Prolog}
\label{conceptualProblem}

\begin{figure}[!h]
\centering
\includegraphics[scale=0.35]{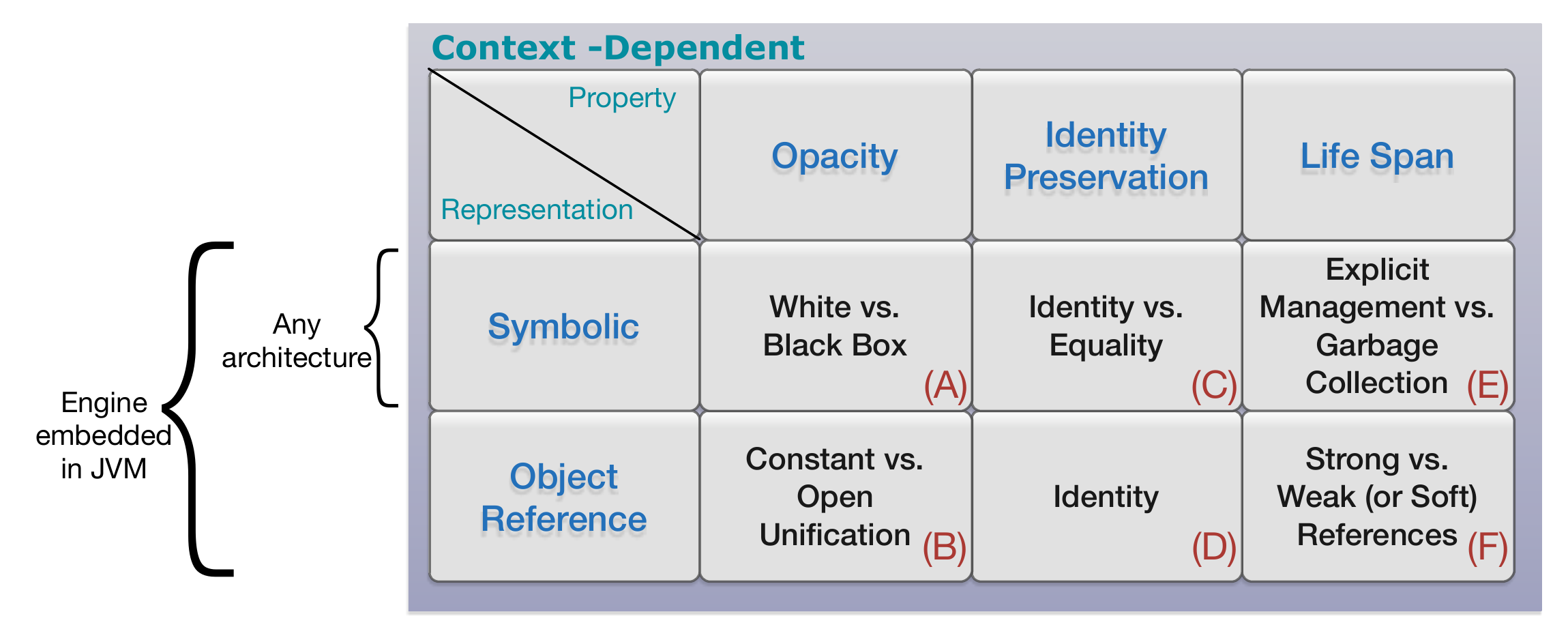}
\caption{Reference Management Dimensions}
\label{fig:ref_dimensions}
\end{figure}

In this section, we identify the different dimensions to be taken into consideration when looking at the problem of representing Java references in Prolog (figure \ref{fig:ref_dimensions}).
These dimensions have been extracted and generalised from existing solutions to this problem both in Prolog and other inter-language representation domains.

\vspace{-0.6em}
\vspace{-1.6em}

\subsection{Reference Representation}
\label{symbolicVsReferences}

A first important dimension is how Java objects are represented on the logic side.
Several integration libraries allow to reify Java objects in Prolog using a symbolic term representation \cite{jpl,Sicstus:manual95,2014-jelia-calejo}.
%(e.g., 
%\textsc{JPL} \cite{jpl}
%and
%\textsc{InterProlog} \cite{2014-jelia-calejo}, 
%\textsc{Jasper} \cite{Sicstus:manual95} 
%or the \textsc{Ciao Java interface} \cite{ciao-reference-manual-1.14-short}
%). 
%
%Ciao Prolog:
%http://ciao-lang.org/docs/ciao/javart.html#Jesus Correas
% see the definition of the java_object/1 predicate
%Sicstus: 
%https://www.fi.muni.cz/~hanka/sicstus/doc/html/jasper/se/sics/jasper/SICStus.html#newObjectTerm(java.lang.Object)
%This is also supported out-of-the-box by Prolog engines embedded in Java.
As show in figure \ref{fig:ref_dimensions}, such approach has the advantage of not relying on any specific Prolog engine architecture.
%In addition, a programmer can decide on the best term reification of an arbitrary Java object without relying on any special feature of the Prolog engine.

Alternatively, Prolog implementations running in the JVM may support the storage of direct object references 
%(i.e., the object itself, not a symbolic representation of it)
(e.g., Jinni \cite{conf/iclp/Tarau04} and LeanProlog \cite{ismm11tarau}).
% TODO find out if TuProlog should also be here
An advantage of this representation scheme is that there are no performance penalties associated to the marshalling/unmarshalling of Java objects to/from the Prolog engine.
% SHOULD WE SAY THAT IN THE OTHER HAND THE OVERALL PERFORMANCE OF THE ENGINE IS TYPICALLY WORSE THAN THOSE OF NATIVE ENGINES ?

%The rest of the dimensions presented in this section are relevant to the symbolic representation (i.e., by means of a Prolog term) of external Java references in Prolog.
%This is because most of these problems are trivially solved in Prolog engines embedded in the JVM.

\vspace{-1em}
\subsection{Opacity of the Representation}
\label{opacityRepresentation}

A second important dimension is the degree of opacity of the representation (i.e., the degree of data exposed). % when reifying a Java object reference in Prolog.
For symbolic term representations \emph{(A)}, frequently a fine-grained reification of the internal object structure (i.e., a white box representation) is desired. For example, \textsc{JTransformer} \cite{Kniesel:2007kx} allows to reason over the structure of terms reifying objects modelling a Java abstract syntax tree.
However, if inspecting the object's structure on the Prolog side is not required, having an opaque reference (i.e., a black box representation) to the corresponding Java object is preferable (e.g., an opaque reference to a GUI component on the Java side).
%This requires an automatic mechanism to generate such opaque term representations of Java objects.
In those cases, an automatic mechanism to generate opaque term representations of Java objects is desirable. 
%Otherwise, the programmer would have the burden to ensure the uniqueness of an arbitrarily chosen black box representation.

When the Prolog engine is embedded in a JVM, a more direct kind of reference to Java objects can be established \emph{(B)}.
In the simplest case, the object reference can be considered and unified as a special constant term.
In spite of the more direct mapping (no automated mapping to generate the reference is required; the term wraps the object `as is'), this case is conceptually equivalent to mapping the object reference to an opaque term representation.
But we may want to combine the best of both worlds and have direct references to the actual Java objects, while still allowing Prolog programs to reason over the internal structure of such objects. 
Approaches such as \textsc{Soul} \cite{Roover:2011uq} have achieved this through the mechanism of \emph{open unification} \cite{2007:sccc:brichau}.
This approach consists in allowing the programmer to customise not the term representation of an object, but rather its unification mechanism.
In a nutshell, the unification mechanism is opened up so that Java objects are not regarded as constants but can be unified with structured logic terms of the right form. 
%Although conceptually different, from several practical aspects this is similar to mapping the object reference to a white box term representation, with the notable difference that an actual reference to the original Java object is kept.

\vspace{-1em}
%not happy with this title
\subsection{Object Identity Preservation}
\label{preservingReferences}

For logic engines running in the JVM \emph{(D)} 
%there is no problem with preserving the identity of the Java objects referred to on the logic side. 
object references are preserved automatically since the term wraps the object `as is'.
For engines not embedded in the JVM, a programmer needs to decide if an object reified as a term should preserve its identity when the term is translated back to a Java object \emph{(C)}.
In many situations, it is not important to preserve such identity (e.g., instances of \lstinline{String}) and a different reference, considered equivalent to the original object (e.g., by means of the \lstinline{equals} method), is acceptable.
%is an acceptable outcome.
However, in certain cases, keeping track of the original reference is required to guarantee the expected behaviour of the program (e.g., if the reference points to a GUI component).
Furthermore, passing around symbolic representations of object references is often more efficient than marshalling and unmarshalling large Java objects.
Note that the need for preserving the original object identity is orthogonal to the required opacity of the representation.
I.e., independently if the reference should be preserved or not, the programmer should still be able to decide on the best representation of the object on the Prolog side.

\vspace{-1em}
\subsection{Reference Life Span}
\label{referenceLifeSpan}

A fourth dimension is the life span of Prolog references to Java objects. 
For a symbolic term representation, a programmer should decide on a mechanism for delimiting the life span of a mapping between a 
Java reference and a Prolog term \emph{(E)}.
This mechanism can be explicit (e.g., an API allowing to request to `forget' a mapping) or rely on JVM garbage collection mechanisms.
An explicit mechanism enables a fine-grained control over the life span of a reference.
For example, a symbolic term representation of an object that is not explicitly referenced in a program (i.e., normally to be scheduled for garbage collection) can still remain valid until explicitly discarded.
Alternatively, a reference life span may be automatically delimited by the JVM garbage-collection mechanism  (e.g., a reference to the application main window). For an object reference representation \emph{(F)}, the programmer may want to keep the reference alive as long as it is present in the Prolog database (i.e., a strong reference).
However, in certain scenarios a Java reference stored in Prolog should not prevent it from being garbage collected (e.g., the reference points to a disposed GUI component).
In that case, the reference should be invalidated when it is reclaimed by the garbage collector.
%For both symbolic and object reference representations, 
A programmer may also want to define customisable cleaning tasks to be automatically executed when a reference is garbage collected.
For example, clauses containing dead references may be automatically retracted from the Prolog database to avoid unexpected behaviours (e.g., null pointer exceptions).
Furthermore, references that may be reclaimed by the garbage collector should be classified according to the  Java (garbage-collected) reference types:
\emph{Weak} for eagerly collected references (discarded at the next garbage collection cycle) and \emph{Soft} for references not aggressively reclaimed (only collected when the memory is tight).\footnote{\url{http://docs.oracle.com/javase/7/docs/api/java/lang/ref/Reference.html}}

\vspace{-1em}
\subsection{Scope of the Inter-Language Conversion Policies}
\label{scopeRepresentation}

%In addition to deciding which of the above choices are most appropriate for a specific programming task, 
We claim that it is useful for a programmer to be able to choose different reference management policies in different parts of the program. To achieve that, it is needed a simple mechanism for scoping and encapsulating the best reference handling policy for certain objects.
Besides greater flexibility, this facilitates performance tuning and testing (e.g., generating mocking representations of references). Next, we will introduce the architecture of a library that supports a customisable management of all these dimensions.

\vspace{-1em}

\section{Architecture}
\label{architecture}

\begin{figure}[!h]
\centering
\includegraphics[scale=0.33]{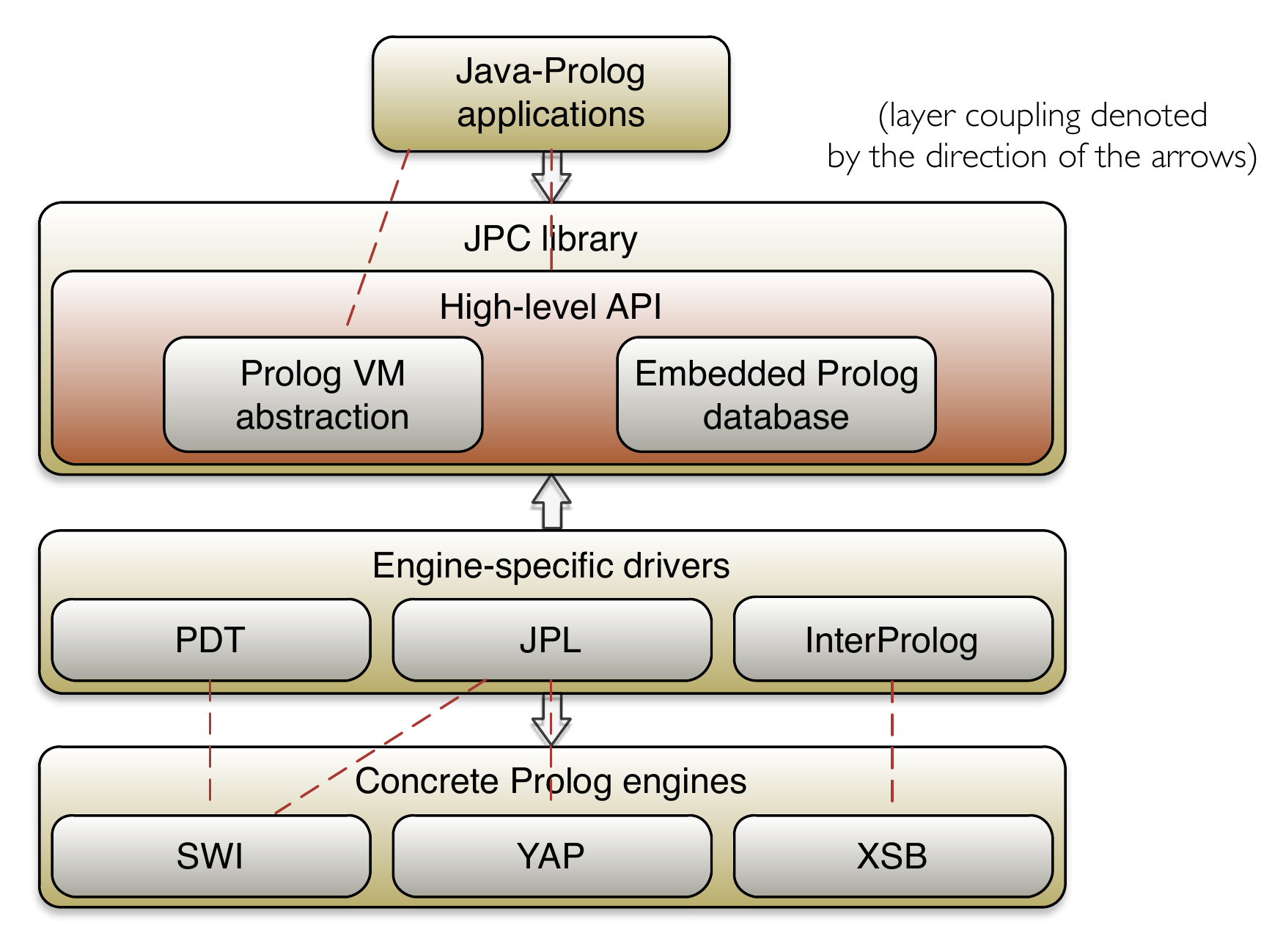}
\caption{The JPC architecture}
\label{fig:jpc_architecture}
\end{figure}

%\subsection{Java--Prolog Applications}
\textsc{JPC} is an integration library supporting the development of hybrid Java--Prolog programs. It provides different levels of abstractions, simplifying the implementation of common inter-operability tasks. To set the ground for discussing the \textsc{JPC} features for Java reference management in Prolog, this section overviews its main components (figure \ref{fig:jpc_architecture}).

%At the top level of figure \ref{fig:jpc_architecture} we have hybrid applications consisting of both Java and Prolog components interacting by means of our library.
%From the user perspective, these applications can be launched either as a Java program with embedded Prolog modules, or as a Prolog program with embedded Java components.
%
%In the rest of this section we discuss the main \textsc{JPC} components.

\vspace{-1em}
\subsection{Prolog VM Abstraction}
\label{prologVMAbstraction}

Several integration libraries rely on the notion of a Prolog engine as a convenient abstraction for interacting with a Prolog virtual machine from Java
%~\cite{Tarau:2012:BEA:2139873.2139878,conf/iclp/Tarau04,pdt,2014-jelia-calejo}.
~\cite{conf/iclp/Tarau04,pdt,2014-jelia-calejo}.
%Quoting Tarau [14]: \emph{``An engine is a language processor reflected through an API that allows its computations to be controlled interactively''}. Tarau defines a logic engine as \emph{``an engine running a Horn Clause Interpreter with LD-resolution~\cite{tarau1993nonstandard} on a given clause database, together with a set of built-in operations''}.
%FROM TARAU PAPER: It has been a long tradition of logic programming languages [20.21] to use multiple logic engines for supporting concurrent execution.
%
%Assuming a Prolog engine implementing most ISO and defacto standard Prolog predicates, we require that such ...
In \textsc{JPC}, a programmer interacts with a Prolog engine abstraction that communicates with concrete Prolog engines using drivers. With portability in mind, when modelling such an abstract Prolog engine we tried to find a compromise between 
(1) offering convenient features facilitating the interaction from Java programs 
and (2) not assuming a specific implementation architecture of the underlying Prolog engine.
%and (3) not assuming non-standard features that may not be available in all Prolog implementations that we envisage to support.
%
Our Prolog engine abstraction provides a general purpose API for interacting with Prolog. However, as illustrated in section \ref{jpc}, \textsc{JPC} also supplies a higher level API that simplifies certain tasks (e.g., inter-language conversions). % and reduces the amount of boilerplate code.
%As part of our abstract Prolog engine, 
\textsc{JPC} defines a set of classes reifying Prolog data types: \lstinline{Term}, \lstinline{Atom}, \lstinline{Compound}, \lstinline{IntegerTerm}, \lstinline{FloatTerm}, \lstinline{Var}, \lstinline{JRef} (a Java reference term; a special kind of term wrapping a Java reference).%, and \lstinline{Query} (a reification of a Prolog query).

%\smallskip
%
%\begin{description}[topsep=0pt,parsep=0pt,itemsep=0pt]
%    \item[Term]: An abstract Prolog term.
%    \item[Atom]: A sequence of characters representing a Prolog atom.
%    \item[Compound]: A compound term consisting of a name and a list of arguments.
%    \item[IntegerTerm]: A Prolog integer term.
%    \item[FloatTerm]: A Prolog float term.
%    \item[Var]: A Prolog variable.
%    \item[JRef]: A Java reference term. A special kind of term wrapping a Java reference.
%    \item[Query]: The reification of a Prolog query. % with convenient methods for obtaining its solutions, where each solution is a map of Prolog variable names to terms.
%\end{description}
%
%\smallskip

\noindent
\vspace{-0.6em}
\vspace{-1.6em}

\subsection{Embedded Prolog Database}
\label{embeddedJPC}

JPC uses an embedded Prolog database running on the JVM and supporting the storage of Java object references in addition to standard Prolog terms.
Several \textsc{JPC} interoperability features rely on this component, which maintains mappings between Prolog terms and arbitrary Java objects (represented as \lstinline{JRef} terms).

\vspace{-1em}

\section{Reference Management with \textsc{JPC}}
\label{jpc}

%This section describes JPC's support for the different reference management dimensions showed in figure \ref{fig:ref_dimensions}.
This section describes JPC's support for the different dimensions related to the management of Java references in Prolog (figure \ref{fig:ref_dimensions}).
%Given space constraints, we make use in most of our examples of the high level \textsc{JPC} API only.% (e.g., transparent conversion between Java objects and terms by means of a conversion context).

\vspace{-1em}
\subsection{Symbolic Representation}
\label{symbolicRepresentation}

To illustrate the properties of symbolic references (identified by the first row of figure \ref{fig:ref_dimensions}), we start by defining a \lstinline{Person} class (listing \ref{lst:personClass}) declaring \lstinline{name} as its only instance variable.

\begin{lstlisting}[language=Java, captionpos=b, frame=lines, rulesepcolor=\color{gray}, numbers=left, numberstyle=\tiny\color{blue}, showspaces=false, showstringspaces=false, breaklines=true, breakatwhitespace=true, caption=The \lstinline{Person} class, label=lst:personClass]
public  class Person implements Serializable {
	private final String name;
	public Person(String name) {this.name = name;}
	...
	@Override
	public boolean equals(Object obj) {
		... return ((Person)obj).name.equals(name); //simplified implementation
	}	
}
\end{lstlisting}

The \lstinline{PersonConverter} class (listing \ref{lst:personConverterClass}) defines how instances of class \lstinline{Person} are translated to a Prolog compound term (lines 5--7) and back (lines 8--10).
According to our classification in section \ref{opacityRepresentation}, the term reification of a person, according to this converter, corresponds to a white box representation since it exposes its internal data.

\begin{lstlisting}[language=Java, captionpos=b, frame=lines, rulesepcolor=\color{gray}, numbers=left, numberstyle=\tiny\color{blue}, showspaces=false, showstringspaces=false, breaklines=true, breakatwhitespace=true, caption=The \lstinline{PersonConverter} class, label=lst:personConverterClass]
public class PersonConverter implements FromTermConverter<Compound, Person>, 
		ToTermConverter<Person, Compound> {
	public static final String PERSON_FUNCTOR_NAME = "person";
	
	@Override public Compound toTerm(Person person, Class<Compound> termClass, Jpc context) {
		return new Compound(PERSON_FUNCTOR_NAME, asList(new Atom(person.getName())));
	}
	@Override public Person fromTerm(Compound personTerm, Type targetType, Jpc context) {
		return new Person(((Atom)((Compound)personTerm).arg(1)).getName());
	}
}
\end{lstlisting}

%\ref{lst:testWhiteBoxAndEqualityLowLevel}

%\begin{lstlisting}[language=Java, basicstyle=\tiny, captionpos=b, frame=lines, rulesepcolor=\color{gray}, numbers=left, numberstyle=\tiny\color{blue}, showspaces=false, showstringspaces=false, breaklines=true, breakatwhitespace=true, caption=White Box without Identity Preservation (low level API), label=lst:testWhiteBoxAndEqualityLowLevel]
%@Test
%final String STUDENT_FUNCTOR_NAME = "student";
%PrologEngine prologEngine = getPrologEngine();
%Person person = new Person("Mary");
%prologEngine.assertz(new Compound(STUDENT_FUNCTOR_NAME, asList(person.toTerm())));
%Query query = prologEngine.query(new Compound(STUDENT_FUNCTOR_NAME, asList(new Var("Person"))));
%Solution solution = query.oneSolutionOrThrow();
%Term personTerm = solution.get("Person");
%Person queriedPerson = Person.create(personTerm);
%assertEquals(person, queriedPerson);
%assertFalse(person == queriedPerson);
%\end{lstlisting}

%Using this converter, 
Listing \ref{lst:testWhiteBoxAndEquality} illustrates a white box term representation of a Java object, without object identity preservation (the first three lines are common to most examples; we will not repeat them).
A central artefact in our approach is a \emph{conversion context}, instantiated in line 4 using a builder class and configured with the \lstinline{PersonConverter} converter.
%As the previous lines, the \lstinline{ctx} variable always refer to this context unless explicitly stated the contrary. 
With this context we obtain the conversion of a person in line 5 (\lstinline{person(mary)}).
Next, we assert the fact \lstinline{student(person(mary))} (line 6).
A \lstinline{student(A)} goal is instantiated in line 7 passing the context defined before.
A person is queried in line 8 using a deterministic query. 
The \lstinline{selectObject()} method adapts each solution to the query as an object whose term reification is given as a string.
This adaptation corresponds to the conversion as a Java object of the term that has been bound to the  \lstinline{Person} variable in the solution.
Lines 9 and 10 verify that the queried and the original persons are equal, although with different identities.

\begin{lstlisting}[language=Java, captionpos=b, frame=lines, rulesepcolor=\color{gray}, numbers=left, numberstyle=\tiny\color{blue}, showspaces=false, showstringspaces=false, breaklines=true, breakatwhitespace=true, caption=White Box without Identity Preservation, label=lst:testWhiteBoxAndEquality]
final String STUDENT_FUNCTOR_NAME = "student";
PrologEngine prologEngine = getPrologEngine();
Person mary = new Person("Mary");
Jpc ctx = JpcBuilder.create().register(new PersonConverter()).build();
Term personTerm = ctx.toTerm(mary);
prologEngine.assertz(new Compound(STUDENT_FUNCTOR_NAME, asList(personTerm)));
Query query = prologEngine.query(new Compound(STUDENT_FUNCTOR_NAME, asList(new Var("Person"))), ctx);
Person queriedPerson = query.<Person>selectObject("Person").oneSolutionOrThrow();
assertEquals(mary, queriedPerson);
assertFalse(mary == queriedPerson);
\end{lstlisting}

Listing \ref{lst:testWhiteBoxAndIdentity} illustrates the mapping of a reference to a term representation (line 2) in the scope of a context.
The \lstinline{newRefTerm()} method associates a person reference (first argument) to an arbitrary (compound) term representation (second argument).
In this example, the term corresponds to the term conversion of the reference according to a given conversion context (obtained by the \lstinline{toTerm()} method of the context instance).
We verify that this time the queried person corresponds to the original person reference in line 6.

\begin{lstlisting}[language=Java, captionpos=b, frame=lines, rulesepcolor=\color{gray}, numbers=left, numberstyle=\tiny\color{blue}, showspaces=false, showstringspaces=false, breaklines=true, breakatwhitespace=true, caption=White Box and Identity Preservation, label=lst:testWhiteBoxAndIdentity]
Jpc ctx = JpcBuilder.create().register(new PersonConverter()).build();
Term personTerm = ctx.newRefTerm(person, ctx.<Compound>toTerm(mary));
prologEngine.assertz(new Compound(STUDENT_FUNCTOR_NAME, asList(personTerm)));
Query query = prologEngine.query(new Compound(STUDENT_FUNCTOR_NAME, asList(new Var("Person"))), ctx);
Person queriedPerson = query.<Person>selectObject("Person").oneSolutionOrThrow();
assertTrue(mary == queriedPerson);
\end{lstlisting}

An example of a black box representation is shown in listing \ref{lst:testBlackBoxAndEqualityBySerialization}.
Here, we assert a term of the form \lstinline{student(serialisation)}, where the compound argument corresponds to the term representation of the serialisation of a \lstinline{Person} instance.
No converter is passed to the query in line 2.
This is because the default conversion context (employed by the query if no context is explicitly passed) includes a converter able to deserialize a Java object from the term representation of its serialisation.
Finally, we verify that our queried person is equal to the original person (line 4) although having different identities (line 5).

Although in the context of this example we have presented this term reification as a black box representation, note that in other contexts this may be considered as a white box.
This would be the case if the Prolog side is intended to interpret such representation (e.g., if it reasons over the serialised bytes of the object \cite{2014-jelia-calejo}).

\begin{lstlisting}[language=Java, captionpos=b, frame=lines, rulesepcolor=\color{gray}, numbers=left, numberstyle=\tiny\color{blue}, showspaces=false, showstringspaces=false, breaklines=true, breakatwhitespace=true, caption=Black Box without Identity Preservation, label=lst:testBlackBoxAndEqualityBySerialization]
prologEngine.assertz(new Compound(STUDENT_FUNCTOR_NAME, asList(SerializedTerm.serialize(mary))));
Query query = prologEngine.query(new Compound(STUDENT_FUNCTOR_NAME, asList(new Var("Person"))));
Person queriedPerson = query.<Person>selectObject("Person").oneSolutionOrThrow();
assertEquals(mary, queriedPerson);
assertFalse(mary == queriedPerson);
\end{lstlisting}

%
%A programmer can also associate an arbitrary term, without any possibly semantic meaning, to a reference (listing \ref{lst:testBlackBoxAndIdentityAdHoc}).
%Therefore, no converter is required when instantiating the context (line 1).
%Instead, we make use again of the \lstinline{newRefTerm()} method to associate this time a person to the arbitrary term \lstinline{cool_student(42)}.
%The same as listing \ref{lst:testWhiteBoxAndIdentity}, the original and queried persons correspond to the same reference (line 6).

%\begin{lstlisting}[language=Java, basicstyle=\tiny, captionpos=b, frame=lines, rulesepcolor=\color{gray}, numbers=left, numberstyle=\tiny\color{blue}, showspaces=false, showstringspaces=false, breaklines=true, breakatwhitespace=true, caption=Black Box and Identity Preservation (arbitrary representation), label=lst:testBlackBoxAndIdentityAdHoc]
%Jpc ctx = JpcBuilder.create().build();
%Term personTerm = ctx.newRefTerm(mary, new Compound("cool_student", asList(new IntegerTerm(42))));
%prologEngine.assertz(new Compound(STUDENT_FUNCTOR_NAME, asList(personTerm)));
%Query query = prologEngine.query(new Compound(STUDENT_FUNCTOR_NAME, asList(new Var("Person"))), ctx);
%Person queriedPerson = query.<Person>selectObject("Person").oneSolutionOrThrow();
%assertTrue(mary == queriedPerson);
%\end{lstlisting}

%A disadvantage of the previous approach is that the programmer is forced to ensure the uniqueness of the terms associated with arbitrary object references.
%This issue is transparently managed by our library.
%

A programmer can also associate an automatically generated term to a reference.
An example is given in listing \ref{lst:testBlackBoxAndIdentityGenerated}.
This time we invoke the method \lstinline{newRefTerm()} passing as only argument the reference to reify as a term (line 2). A (black box) term representation is generated behind the curtains.
Our library guarantees that such generated term representations are identical for the same object even across different contexts.

\begin{lstlisting}[language=Java, captionpos=b, frame=lines, rulesepcolor=\color{gray}, numbers=left, numberstyle=\tiny\color{blue}, showspaces=false, showstringspaces=false, breaklines=true, breakatwhitespace=true, caption=Black Box and Identity Preservation, label=lst:testBlackBoxAndIdentityGenerated]
Jpc ctx = JpcBuilder.create().build();
Term personTerm = ctx.newRefTerm(mary);
prologEngine.assertz(new Compound(STUDENT_FUNCTOR_NAME, asList(personTerm)));
Query query = prologEngine.query(new Compound(STUDENT_FUNCTOR_NAME, asList(new Var("Person"))), ctx);
Person queriedPerson = query.<Person>selectObject("Person").oneSolutionOrThrow();
assertTrue(mary == queriedPerson);
\end{lstlisting}

As discussed in section \ref{referenceLifeSpan}, a programmer should also be able to control the life span of term--reference mappings. Listing \ref{lst:testExplicitManagementLifeSpan} shows an example.
%As in listing \ref{lst:testWhiteBoxAndIdentity}, 
We use the \lstinline{newRefTerm()} method (line 2) to associate a reference to its (context dependent) term reification.
But afterwards we delete this association using the \lstinline{forgetRefTerm()} method (line 5).
Thus, although the queried person is equal to the original person (line 7) since the term is translated according to the conversion context (line 1), they do not have the same identity (line 8) as the association between the term and the original reference was eliminated.

\begin{lstlisting}[language=Java, captionpos=b, frame=lines, rulesepcolor=\color{gray}, numbers=left, numberstyle=\tiny\color{blue}, showspaces=false, showstringspaces=false, breaklines=true, breakatwhitespace=true, caption=Explicit Management of Associations Life Span, label=lst:testExplicitManagementLifeSpan]
Jpc ctx = JpcBuilder.create().register(new PersonConverter()).build();
Term personTerm = ctx.newRefTerm(person, ctx.<Compound>toTerm(mary));
prologEngine.assertz(new Compound(STUDENT_FUNCTOR_NAME, asList(personTerm)));
assertTrue(mary == prologEngine.query(new Compound(STUDENT_FUNCTOR_NAME, asList(new Var("Person"))), ctx).selectObject("Person").oneSolutionOrThrow());
ctx.forgetRefTerm((Compound)personTerm);
Person queriedPerson = prologEngine.query(new Compound(STUDENT_FUNCTOR_NAME, asList(new Var("Person"))), ctx).<Person>selectObject("Person").oneSolutionOrThrow();
assertEquals(mary, queriedPerson);
assertFalse(mary == queriedPerson);
\end{lstlisting}

A programmer can also rely on the Java garbage collection mechanism for delimiting the life span of an association as shown in listing \ref{lst:testGarbageCollectionLifeSpan}.
The \lstinline{newWeakRefTerm()} method (line 2) is equivalent to the \lstinline{newRefTerm()} method discussed earlier.
But in this case the association between a term and a reference persists as long as the reference is not reclaimed in the next garbage collection cycle.
To prove it, we assign \lstinline{null} to the only variable keeping a reference to the person (line 4) and give a hint to the garbage collector to start a cycle (line 5).
Note that the query is not instantiated with a conversion context (line 7).
Thus, an exception is raised when we try to convert the term (bound to the variable \lstinline{Person}) to an object as no converter is found and no reference is associated to such term.
Our framework also provides the \lstinline{newSoftRefTerm()} method with similar semantics than \lstinline{newWeakRefTerm()}, with the only difference that an association between a term and a reference may persist some time after a garbage collection cycle, and will be deleted only if the memory gets tight.

\begin{lstlisting}[language=Java, captionpos=b, frame=lines, rulesepcolor=\color{gray}, numbers=left, numberstyle=\tiny\color{blue}, showspaces=false, showstringspaces=false, breaklines=true, breakatwhitespace=true, caption=Garbage Collection Management of Associations Life Span, label=lst:testGarbageCollectionLifeSpan]
Jpc ctx = JpcBuilder.create().register(new PersonConverter()).build();
Term personTerm = ctx.newWeakRefTerm(mary, ctx.<Compound>toTerm(mary));
prologEngine.assertz(new Compound(STUDENT_FUNCTOR_NAME, asList(personTerm)));
mary = null;
System.gc();
try {
	prologEngine.query(new Compound(STUDENT_FUNCTOR_NAME, asList(new Var("Person")))).<Person>selectObject("Person").oneSolutionOrThrow();
	fail();
} catch(ConversionException e) {}
\end{lstlisting}

\vspace{-1em}
\subsection{Object Reference Representation}
\label{objectReferenceRepresentation}

This section focuses on the properties of object references (identified by the second row of figure \ref{fig:ref_dimensions}).
Although our library currently only has drivers for non-embedded Prolog engines, as a proof of concept we implement the examples in this section using the \textsc{JPC} embedded Prolog database described in section \ref{prologVMAbstraction}.
With the exception of open unification, all the other properties are supported by our implementation.

We start with an example of constant unification of references in listing \ref{lst:testConstantUnification}.
As mentioned in section \ref{prologVMAbstraction}, a \textsc{JPC}\lstinline{JRef} term wrapps an object reference.
In our current version, they are unified as constants (i.e., unifying tow \lstinline{JRef} terms succeeds if their referred objects are equal). In line 1 we assert that \emph{mary} (wrapped in a \lstinline{JRef} term) is a student. In line 2 we query if a different person object with the same name is a student, which succeeds.

\begin{lstlisting}[language=Java, captionpos=b, frame=lines, rulesepcolor=\color{gray}, numbers=left, numberstyle=\tiny\color{blue}, showspaces=false, showstringspaces=false, breaklines=true, breakatwhitespace=true, caption=Constant Unification of \lstinline{JRef} terms, label=lst:testConstantUnification]
prologEngine.assertz(new Compound(STUDENT_FUNCTOR_NAME, asList(JRef.jRef(mary))));
assertTrue(prologEngine.query(new Compound(STUDENT_FUNCTOR_NAME, asList(JRef.jRef(new Person("mary"))))).hasSolution());
Solution solution = prologEngine.query(new Compound(STUDENT_FUNCTOR_NAME, asList(new Var("X")))).oneSolutionOrThrow();
JRef<Person> jRef = (JRef<Person>) solution.get("X");
assertTrue(mary == jRef.getReferent());
\end{lstlisting}

Thanks to our embedded Prolog database, the identity of a reference is trivially preserved.
To illustrate this, we execute a deterministic query (line 3) with goal \lstinline{student(X)}.
%The query solution, an instance of the \lstinline{Solution} class, exposes a map interface where keys correspond to unbound variables in the goal and the values to the bound terms in the solution.
We verify that the obtained referent has the same identity as \emph{mary} in line 5.

%Since we are using an embedded Prolog database, the identity of a reference is trivially preserved, as illustrated in listing \ref{lst:testIdentityPreservation}.
%As the previous example, we assert first that \emph{mary} is a student (line 1).
%Afterwards, we execute a deterministic query (line 2) to obtain the solution to the query.
%Such solution is an instance of the \lstinline{Solution} class, that exposes a map interface where keys correspond to unbound variables in the query and the values to the bound terms according to a solution.
%We verify that the obtained referent has the same identity than \emph{mary} in line 4.

%\begin{lstlisting}[language=Java, basicstyle=\tiny, captionpos=b, frame=lines, rulesepcolor=\color{gray}, numbers=left, numberstyle=\tiny\color{blue}, showspaces=false, showstringspaces=false, breaklines=true, breakatwhitespace=true, caption=Identity Preservation in \lstinline{JRef} terms, label=lst:testIdentityPreservation]
%prologEngine.assertz(new Compound(STUDENT_FUNCTOR_NAME, asList(JRef.jRef(mary))));
%Solution solution = prologEngine.query(new Compound(STUDENT_FUNCTOR_NAME, asList(new Var("X")))).oneSolutionOrThrow();
%JRef<Person> jRef = (JRef<Person>) solution.get("X");
%assertTrue(mary == jRef.getReferent());
%\end{lstlisting}

Listing \ref{lst:testLifeSpan} shows how to create \lstinline{JRef} instances that may be garbage collected.
 We first create two objects equal to \emph{mary} and assert them, using two kind of references: \emph{strong} (line 3) and \emph{weak} (line 4). % and \emph{soft} (line 5). 
When we query for students unifying with \emph{mary} (line 5) using a strong reference, we get two results instead of one.
This is because the unification semantics of \lstinline{JRef} terms evaluates the referents, not the actual \lstinline{JRef} term wrapper.
%Listing \ref{lst:testLifeSpan} also shows an example of an invalidated reference.
Afterwards we assign to \lstinline{null} the variable \lstinline{person2} (line 6) and give a hint to the garbage collector to execute a cycle (line 7).
Since the referent of the \lstinline{JRef} term asserted in line 4 has been invalidated, the number of students unifying with \emph{mary} is now only 1 (line 8).
Note that weak or soft references should be used with care: they may require non-monotonic reasoning as the referent of a \lstinline{JRef} term may be invalidated during the query execution.

\begin{lstlisting}[language=Java, captionpos=b, frame=lines, rulesepcolor=\color{gray}, numbers=left, numberstyle=\tiny\color{blue}, showspaces=false, showstringspaces=false, breaklines=true, breakatwhitespace=true, caption=Life Span of \lstinline{JRef} terms, label=lst:testLifeSpan]
Person person2 = new Person("Mary");
Person person3 = new Person("Mary");
prologEngine.assertz(new Compound(STUDENT_FUNCTOR_NAME, asList(JRef.jRef(mary))));
prologEngine.assertz(new Compound(STUDENT_FUNCTOR_NAME, asList(JRef.weakJRef(person2))));
assertEquals(2, prologEngine.query(new Compound(STUDENT_FUNCTOR_NAME, asList(JRef.jRef(mary)))).allSolutions().size());
person2 = null;
System.gc();
assertEquals(1, prologEngine.query(new Compound(STUDENT_FUNCTOR_NAME, asList(JRef.jRef(mary)))).allSolutions().size());
\end{lstlisting}

The previous example motivates the need of a cleaning mechanism.
Listing \ref{lst:testCleaningTask} illustrates such mechanism using a user-defined cleaning task.
To keep our example simple, this cleaning task retracts all the asserted students (lines 1--5) when a reference is invalidated. A more sophisticated example would retract only the invalidated reference.
Our cleaning task is associated with a weak reference in line 6.
In line 9 we verify that no students are in the database after the reference to \emph{mary} has been invalidated (lines 7--8).

\begin{lstlisting}[language=Java, captionpos=b, frame=lines, rulesepcolor=\color{gray}, numbers=left, numberstyle=\tiny\color{blue}, showspaces=false, showstringspaces=false, breaklines=true, breakatwhitespace=true, caption=Cleaning Tasks, label=lst:testCleaningTask]
Runnable cleaningTask = new Runnable() {
	@Override public void run() {
		prologEngine.retractAll(new Compound(STUDENT_FUNCTOR_NAME, asList(Var.ANONYMOUS_VAR)));
	}
};
prologEngine.assertz(new Compound(STUDENT_FUNCTOR_NAME, asList(JRef.weakJRef(mary, cleaningTask))));
mary = null;
System.gc();
assertFalse(prologEngine.query(new Compound(STUDENT_FUNCTOR_NAME, asList(Var.ANONYMOUS_VAR))).hasSolution());
\end{lstlisting}

\vspace{-1em}

\section{Related Work}
\label{relatedWork}

Most related work has already been overviewed in sections \ref{conceptualProblem} and \ref{prologVMAbstraction} 
so we do not repeat it here.
\textsc{InterProlog} inspired the serialisation mechanism illustrated in listing \ref{lst:testBlackBoxAndEqualityBySerialization}.
%Regarding this conversion approach, \textsc{InterProlog} 
It provides a more structured representation of a serialised object on the Prolog side using a definite clause grammar. Currently we represent serialised bytes as an atom using a raw base-64 encoding. %~\cite{rfc4648}).
\textsc{InterProlog} has limited support, however, for customising the reification as a term of arbitrary Java objects (even not serialisable ones) as in our approach. 
%It also provides predicates for creating (the serialised representation of) new Java objects on the Prolog side using an object template mechanism.
%There is limited support, however, for customising the reification as a term of arbitrary Java objects (even no serialisable ones) as in our approach. 
%A portability limitation of InterProlog is that its current implementation assumes a Prolog engine as being an external program to be invoked, limiting its applicability to other kinds of architectures.
%
%The JPL library is compatible with both SWI-Prolog \cite{wielemaker:2011:tplp} and YAP \cite{Costa:2012fk}. A discussion on the techniques employed for reaching compatibility between SWI-Prolog and YAP libraries is presented in \cite{DBLP:conf/padl/WielemakerC11}.
%In JPC, the objects reifying Prolog terms in the Java side were mainly inspired by JPL.
%As InterProlog, JPL also makes strong assumptions about the underlying Prolog engine, which may compromise portability. Particularly, that a Prolog engine is connected by means of the JNI library.
%The PDT Connector library, although it is currently compatible with SWI only, plans to be ported to YAP in the near future. 
%To the best of our knowledge, there are no other libraries attempting to solve the problem of Java--Prolog interoperability in a portable way.
Concerning our mechanisms for custom two-way conversions between inter-language artefacts, this was inspired by Google's \textsc{Gson} library, which aims to provide a high-level tool for conversions between Java objects and their JSON representation.
%In fact, certain aspects of our library can be regarded as a re-implementation of \textsc{Gson} for the domain of Java--Prolog artefact conversions.

%\input{futureWork}
\vspace{-1em}

\section{Conclusions and Future Work}
\label{conclusions}

This work discusses different dimensions that should be taken into consideration when dealing with Java references in Prolog programs.
These dimensions have been extracted from many sources, including our own experience, a study of existing approaches, and even existing solutions in other domains.
At the moment, \textsc{JPC} does not implement a mechanism for interacting with Java from the Prolog side.
%Particularly, it is lacking a way of activating Java behaviour from Prolog.
In line with our portability goal, we plan to implement our Prolog side API using Logtalk~\cite{pmoura03}, a portable object-oriented layer for Prolog.
%However, we intend to provide for plain Prolog mechanisms that delegate to our portable interoperability API implemented on top of Logtalk.
As in the current Java side API, we expect to prototype a first version by reusing existing bridge libraries.  
We will also continue improving our embedded Prolog database so that it can be released as a stand-alone embedded Prolog engine.
We hope that our work will benefit not only implementors of Java--Prolog integration libraries, but also integrators of similar object-oriented and logic languages.

%\paragraph{Acknowledgements}
%We are grateful to the authors of the bridge libraries used to build our JPC library drivers for their groundwork and for their help.
%We thank G\"unter Kniesel for his valuable feedback during the drafting of this paper.
%Work partially supported by the LEAP project (PTDC/EIA-CCO/112158/\ 2009), the ER\-DF/COMPETE Program and by the FCT project FCOMP-01-0124-FEDER-022701.

\bibliographystyle{acmtrans}
\bibliography{references}

\end{document}